\documentclass[11pt]{article}
\usepackage{amsmath,epsf,amssymb,latexsym,enumerate,cite,shadow,array,color,dsfont}
\usepackage[ps,dvips,matrix,arrow,frame,import,curve,color]{xy}
\usepackage[mathscr]{euscript}
\renewcommand{\theequation}{\thesection.\arabic{equation}}
\newcounter{saveeqn}
\newcommand{\add}{\addtocounter{equation}{1}}
\newcommand{\alpheqn}{\setcounter{saveeqn}{\value{equation}}%
\setcounter{equation}{0}%
\renewcommand{\theequation}{\mbox{\thesection.\arabic{saveeqn}{\alph{equation}}}}}
\newcommand{\reseteqn}{\setcounter{equation}{\value{saveeqn}}%
\renewcommand{\theequation}{\thesection.\arabic{equation}}}

\newif\iffigs\figstrue

\DeclareFontFamily{U}{rsf}{}
\DeclareFontShape{U}{rsf}{m}{n}{
  <5> <6> rsfs5 <7> <8> <9> rsfs7 <10-> rsfs10}{}
\DeclareMathAlphabet\Scr{U}{rsf}{m}{n}

\def\pplogo{\vbox{\kern-\headheight\kern -29pt
\halign{##&##\hfil\cr&{
\ppnumber}\cr\rule{0pt}{2.5ex}&\ppdate\cr}
}}
\makeatletter
\def\ps@firstpage{\ps@empty \def\@oddhead{\hss\pplogo}%
  \let\@evenhead\@oddhead 
}
\def\maketitle{\par
 \begingroup
 \def\thefootnote{\fnsymbol{footnote}}
 \def\@makefnmark{\hbox{$^{\@thefnmark}$\hss}}
 \if@twocolumn
 \twocolumn[\@maketitle]
 \else \newpage
 \global\@topnum\z@ \@maketitle \fi\thispagestyle{firstpage}\@thanks
 \endgroup
 \setcounter{footnote}{0}
 \let\maketitle\relax
 \let\@maketitle\relax
 \gdef\@thanks{}\gdef\@author{}\gdef\@title{}\let\thanks\relax}
\makeatother

\newcommand{\bea}{\begin{eqnarray}}
\newcommand{\eea}{\end{eqnarray}}
\newcommand{\be}{\begin{equation}}
\newcommand{\ee}{\end{equation}}

\def\R{{\mathbb R}}

\def\sA{{\mathscr A}}

\def\sD{{\mathscr D}}
\def\sF{{\mathscr F}}

\def\cC{{\Scr C}}

\def\cL{{\Scr L}}

\def\bbe{{\bar{\bar{\epsilon}}}}
\def\bbc{{\bar{\bar{\chi}}}}

\def\bbl{{\bar{\bar{\lambda}}}}

\def\CN{{\cal N}}
\def\CO{{\cal O}}
\def\CA{{\cal A}}
\def\CL{{\cal L}}
\def\half{\frac{1}{2}}
\def\lsu{\mathfrak{su}}
\def\lso{\mathfrak{so}}

\def\del{\nabla}
\def\eps{\epsilon}
\def\veps{\varepsilon}
\def\d{\partial}
\def\tH{{\tilde H}}
\def\tB{{\tilde B}}

\begin{document}

\begin{titlepage}

\begin{center}
\hfill BRX-TH-6293\\
\hfill NSF-KITP-15-103\\
\vskip 0.8in

{\Large \bf ${\cal N}=(4,4)$ Vector Multiplets on Curved Two-Manifolds}
\\[2ex]

\vskip 0.3in

{ Albion Lawrence${}^{a,b}$ and Masoud Soroush${}^{a,c,d}$}\\

\vskip 0.3in
{\it $^{a}$Martin A. Fisher School of Physics, Brandeis University}\\
{\it Waltham, MA 02453, USA}\\
{\it $^{b}$Kavli Institute for Theoretical Physics, University of California}\\
{\it Santa Barbara, CA 93106, USA}\\
{\it $^{c}$Department of Mathematics, Brandeis University}\\
{\it Waltham, MA 02453, USA}\\
{\it $^{d}$Department of Mathematics, Coppin State University}\\
{\it Baltimore, MD 21216, USA}

\end{center}

\vskip 0.4in

\begin{center} {\bf Abstract} \end{center}
\vskip 0.05in

We study the necessary conditions for preserving ${\cal N}=(4,4)$ supersymmetry on curved 2d backgrounds, following the strategy of Dumitrescu, Festuccia, and Seiberg. We derive the transformation laws and invariant action for off-shell Abelian vector multiplets. An explicit solution of the supersymmetry conditions is found on the round two-sphere.

\vfill

\noindent  April 2015

{
\let\thefootnote\relax
\footnotetext{\noindent albion@brandeis.edu, soroush@brandeis.edu}
}

\end{titlepage}
\tableofcontents
\section{Introduction}

Localization techniques for supersymmetric quantum field theories on curved Euclidean backgrounds have yielded a remarkable set of exact results.  Pestun's pioneering work computed partition functions and Wilson line expectation values for supersymmetric gauge theories on $S^4$ \cite{Pestun:2007rz}. There has since been a host of results for three-dimensional QFTs \cite{Kapustin:2009kz,Kapustin:2010xq,Drukker:2010nc,Drukker:2011zy,Marino:2011eh,Klemm:2012ii,Awata:2012jb,Hirano:2014bia}. (For a review see \cite{Hosomichi:2014hja}.)

In two dimensions, these techniques were used to compute the exact partition function on $S^2$ for ${\cal N}=(2,2)$ gauged linear sigma models (GLSMs) \cite{Benini:2012ui,Doroud:2012xw}. For GLSMs which flow to nonlinear sigma models with Calabi-Yau target spaces, these results were shown to yield the exact 
K\"ahler potential on the moduli space of K\"ahler deformations of the Calabi-Yau \cite{Jockers:2012dk,Gomis:2012wy,Honma:2013hma,Jia:2013foa}. They have also been used to study 2d theories on spaces with boundaries \cite{Hori:2013ika,DiPietro:2015zia}\ and on nonorientable surfaces \cite{Kim:2013ola}. 

The starting point of these calculations is the realization of supersymmetry on curved backgrounds, which involves constraints on the supersymmetry parameters -- for example, they might be Killing or conformal Killing spinors \cite{Pestun:2007rz,Benini:2012ui,Doroud:2012xw}  -- and a deformation of the flat space algebra. A systematic approach to finding these constraints and deformations was initiated for four-dimensional theories in \cite{Festuccia:2011ws,Dumitrescu:2012ha,Klare:2012gn}. In this approach, one considers off-shell supergravity multiplets in the limit that the Planck scale goes to infinity. Demanding that supersymmetry transformations do not transform the background metric and gravitino constrains the spinors; allowed background auxiliary fields parametrize the deformations of the flat-space superalgebra.\footnote{For earlier work applied to $AdS_4$ backgrounds, see \cite{Adams:2011vw} and \cite{Jia:2011hw}.} These constraints have been carefully analyzed in various cases in four \cite{Festuccia:2011ws,Dumitrescu:2012ha,Klare:2012gn,Closset:2013vra,Closset:2014uda,Closset:2013sxa} and two dimensions \cite{Closset:2014pda}, the latter for ${\cal N}=(2,2)$ theories. In four dimensions, localization techniques have been applied to study the dependence of the partition functions of these theories on the geometry of the spacetime manifold \cite{Closset:2013vra,Closset:2014uda}.

${\cal N} = (4,4)$ backgrounds retain a great deal of interest -- see for example \cite{Katz:1999xq,Eguchi:2010ej,Cheng:2012tq,Harvey:2013mda,Pandharipande:2014qoa,Harvey:2014cva,Cheng:2014zpa,Harvey:2014nha,Katz:2014uaa,Cheng:2015kha}.  In this article, we study manifestly ${\cal N}=(4,4)$ supersymmetric quantum field theories on curved two-dimensional backgrounds, following the procedure in \cite{Festuccia:2011ws,Dumitrescu:2012ha}.\footnote{Equivariant elliptic genera for ${\cal N}=(4,4)$ theories have been computed via localization when the worldsheet is a flat torus, by writing the theory in ${\cal N}=(2,2)$ language \cite{Harvey:2014nha}.}

We will focus on theories of Abelian vector multiplets only.  These have an off-shell realization with a finite number of auxiliary fields.  Our eventual goal is to consider ${\cal N}=(4,4)$ gauged linear sigma models which might for example realize $A_n$ ALE spaces via hyperK\"ahler quotients.  The completely off-shell hypermultiplets which transform linearly under gauge symmetries cannot be realized via a finite number of auxiliary fields; partially off-shell multiplets (which close on an off-shell central charge) with finite set of auxiliary fields do exist \cite{West:1990tg}, as well as hypermultiplets which cannot be separately linearly gauged without breaking the $SU(2)$ R-symmetry \cite{West:1990tg,Gates:1988tn,Gates:1995aj}. We found the story for vector multiplets to be sufficiently rich to merit its own treatment.

This paper is organized as follows. In section~\ref{SUSYreview}, we review the "large" ${\cal N}=(4,4)$ superalgebra, as described in \cite{Sevrin:1988ew}.   In section~\ref{GM}, we study the supersymmetry transformations of the field components of the gravity multiplet and derive the necessary conditions for preserving ${\cal N}=4$ rigid supersymmetry on a curved background. In section~\ref{AdSUSY}, we study the resulting constraints on 2d manifolds which admit ${\cal N}=(4,4)$ supersymmetry. In section~\ref{VM}, we derive the supersymmetry transformations of the vector multiplet propagating on curved backgrounds, and show that they transform under the large ${\cal N}=(4,4)$ superalgebra. In section~\ref{Sphere}, we explicitly solve the supersymmetry equations derived in section~\ref{GM} on the two-sphere. Section~\ref{Conc} is devoted to our concluding remarks, and possible directions for future research. Appendix~\ref{Notation} summarizes our notations and conventions, and presents a list of useful identities used throughout the paper; Appendix~\ref{SOfour}\ reviews the isomorphism of  $SO(4)$ and $SU(2)\times SU(2)$.
\section{${\cal N}=(4,4)$ superalgebra}\label{SUSYreview}

${\cal N}=(4,4)$ superconformal algebras come in a one-parameter family \cite{Gastmans:1987up,Sevrin:1988ew}\ of algebras which generically contain the Kac-Moody algebra $\mathfrak{su}(2)\oplus\mathfrak{su}(2)\oplus\mathfrak{u}(1)$, of which the $\mathfrak{su}(2)$ components correspond to an R symmetry. Two particular values of this parameter correspond to well-known versions of $\CN=(4,4)$ algebras: 
\begin{itemize}
\item The "large" $\CN=4$ superconformal algebra with Kac-Moody subalgebra $\mathfrak{so}(4) \equiv \mathfrak{su}(2)^{(+)}\oplus \mathfrak{su}(2)^{(-)}$ and finite subalgebra $\mathfrak{osp}(4|2)$. This corresponds to $\gamma = \half$ in \cite{Sevrin:1988ew}. The labels $(\pm)$ merely distinguish the two $\mathfrak{su}(2)$ factors. 
\item The "small" superconformal algebra with Kac-Moody subalgebra $\mathfrak{su}(2)$ and finite subalgebra $\mathfrak{su}(2|1,1)$. This corresponds to $\gamma \to 1$ or $\gamma \to 0$ in \cite{Sevrin:1988ew}. This is a subalgebra of the large algebra.  In the limit $\gamma \to 1$, the R-symmetry maps to $\mathfrak{su}(2)^{(+)}$; in the limit $\gamma \to 0$, the R-symmetry maps to $\mathfrak{su}(2)^{(-)}$.
\end{itemize}

In all of the theories in \cite{Sevrin:1988ew}, the supercharges $G$ transform as a vector under $\mathfrak{so}(4)$, with index $a = 1, \ldots, 4$.  If we write $\mathfrak{so}(4) = \mathfrak{su}(2)^{(+)} \oplus \mathfrak{su}(2)^{(-)}$, the generators of $\mathfrak{so}(4)$ can be written in terms of generators $\alpha^{(\pm)I}_{ab}$ of the two $\mathfrak{su}(2)$ subalgebras, where $I = 1, \ldots 3$ :
\begin{equation}\label{SO4}
\big[\alpha^{(\pm) I},\alpha^{(\pm) J}\big]=-\varepsilon^{IJK}\alpha^{(\pm) K}\ ,\qquad  \big[\alpha^{(+)I},\alpha^{(-)J}\big]=0\ ,\qquad \big\{\alpha^{(\pm)I},\alpha^{(\pm) J}\big\}=-\frac{1}{2}\delta^{IJ}\ .
\end{equation} 
A convenient representation for $\alpha^{(\pm)I}_{ab}$ is \cite{Sevrin:1988ew}
\begin{equation}\label{basis}
\alpha^{(\pm) I}_{ab}=\pm\frac{1}{2}\big(\delta^{I}_{a}\delta^{4}_{b}-\delta^{I}_{b}\delta^{4}_{a}\big)+\frac{1}{2}\varepsilon_{Iab}\ ,
\end{equation}  
with the understanding that the $\varepsilon_{Iab}$ symbol vanishes if any of its indices is equal to 4. Appendix \ref{SOfour} describes the decomposition of the vector of $\mathfrak{so}(4)$ into a $({\bf 2},{\bf 2})$ of $\mathfrak{su}(2)\oplus\mathfrak{su}(2)$.

To set notation, we quote the chiral (left-moving) part of the ``large" $\CN=(4,4)$ superconformal algebra \cite{Sevrin:1988ew}:
\begin{equation}\label{alg}
\begin{aligned}
& \big[L_m,L_n\big]=(m-n)L_{m+n}+\frac{1}{12}\,c(m^3-m)\delta_{m+n}\ ,\\
&  \big[L_m,\phi_n\big]=\big((d_{\phi}-1)m-n\big)\phi_{m+n}\ ,\\
&  \big\{G_{m}^{a},G_{n}^{b}\big\}=2\delta^{ab}L_{m+n}+\frac{1}{3}\,c\,\delta^{ab}\delta_{m+n}\Big(m^2-\frac{1}{4}\Big)+2(n-m)\big(\alpha^{(+)I}_{ab}A^{(+)I}_{n-m}+\alpha^{(-)I}_{ab}A^{(-)I}_{n+m}\big)\ ,\\
&  \big[A_{m}^{(\pm) I},A^{(\pm) J}_{n}\big]=\varepsilon^{IJK}A^{(\pm) K}_{m+n}-\frac{1}{6}\,c\,m\,\delta^{IJ}\delta_{m+n}\ ,\\
&  \big[A_{m}^{(+)I},A_{n}^{(-)J}\big]=0\ , \\
\end{aligned}
\end{equation}

\begin{equation}
\begin{aligned}
&  \big[A_{m}^{(\pm) I},G_{n}^{a}\big]=\alpha^{(\pm) I}_{ab}\big(G^{b}_{m+n}\mp m\, Q^{b}_{m+n}\big)\ ,\ \ \ \ \ \big[A^{(\pm) I}_{m},Q^{a}_{n}\big]=\alpha^{(\pm) I}_{ab}\,Q^{b}_{m+n}\ ,\\
& \big\{Q^{a}_{m},G^{b}_{n}\big\}=2\big(\alpha^{(+)I}_{ab}A^{(+)I}_{m+n}-\alpha^{(-)I}_{ab}A^{(-)I}_{m+n}\big)+\delta^{ab}\,U_{m+n}\ ,\\
&  \big[U_m,Q^{a}_{n}\big]=0\ ,\qquad  \big[U_m,G^{a}_{n}\big]=m\,Q^{a}_{m+n}\ ,\qquad  \big[U_m,A^{(\pm) I}_{n}\big]= 0\ ,\\
&  \big[U_m,U_n\big]=-\frac{1}{3}\,c\,m\delta_{m+n}\ ,\qquad \big\{Q^{a}_{m},Q^{b}_{n}\big\}=-\frac{1}{3}\,c\,\delta^{ab}\delta_{m+n}\ ,
\end{aligned}
\end{equation}
Here $c$ is the central charge of the superconformal algebra. In (\ref{alg}), $\{L_m,G^{a}_{m},A^{(\pm) I}_{m},Q^{a}_{m},U_m\}$ are generators of the left-moving chiral Virasoro algebra, chiral supersymmetry generators of dimension 3/2, chiral operators of dimension 1, chiral operators of dimension 1/2, and finally a single chiral operator of dimension 2, respectively. The lower alphabetical indices $m$ are mode numbers for the currents; in the usual fashion, for a chiral operator of dimension $d_{\phi}$, 
\be
	{\cal O}(z) = \sum_n \CO_n z^{-d_{\phi} - n}
\ee
The bosonic operators $A^{(\pm) I}_{n}$ are generators of the $\mathfrak{su}(2)^{(+)}\oplus\mathfrak{su}(2)^{(-)} \equiv \mathfrak{so}(4)$ Kac-Moody R-symmetry.  The indices $(\pm)$ refer to the two $\mathfrak{su}(2)$ factors; index $I$ runs over $\{1,2,3\}$, labeling $\mathfrak{su}(2)$ generators. The fermionic generators $G^{a}_{n}$, and $Q^{a}_{n}$ transform as vectors of the global part of the $\mathfrak{so}(4)$ R-symmetry, where $a\in\{1,2,3,4\}$. $\phi_n$ is a collective notation for the fields $G^{a}_{n}$, $A^{(\pm) I}_{m}$, $Q^{a}_{n}$, and $U_n$, with the corresponding dimension $d_{\phi}$. 

Finally, in the full $\CN=(4,4)$ algebra, there is an equivalent antichiral (right-moving) algebra with operators $\{{\tilde L}_m, {\tilde G}^a_m, {\tilde A}^{(\pm)I}_m,{\tilde Q}^a_m,{\tilde U}_m\}$.

In this work the large ${\cal N}=(4,4)$ theory will play an important role. Following \cite{Benini:2012ui,Doroud:2012xw,Jockers:2012dk}, we imagine studying gauged linear sigma models which flow to nontrivial superconformal theories in the IR; only the finite subalgebra is realized at intermediate points of the renormalization group flow, and this is all that is required for localization. The chiral part of this algebra is generated by operators $\{L_0, L_{\pm 1}, G^a_{\pm 1/2}, A^{(\pm),I}_0\}$, and has the commutation relations:
\begin{equation}\label{finitealg}
\begin{aligned}
& \big[L_0,L_{\pm1}\big]=\mp L_{\pm1}\ , && \big[L_1,L_{-1}\big]=2L_0\ ,\\
& \big[L_0,G^{a}_{\pm 1/2}\big]=\mp\frac{1}{2}G^{a}_{\pm1/2}\ , && \big[L_1,G^{a}_{+1/2}\big]=0\ ,\\
& \big[L_1,G^{a}_{-1/2}\big]=G^{a}_{+1/2}\ , && \big[L_{-1},G^{a}_{+1/2}\big]=-G^{a}_{-1/2}\ ,\\
& \big[L_{-1},G^{a}_{-1/2}\big]=0\ , && \big[L_{0},A^{(\pm) I}_{0}\big]=0\ ,\\
& \big[L_{\pm1},A^{(\pm) I}_{0}\big]=0\ , && \big[A^{(+)I}_{0},A^{(-)J}_{0}\big]=0\ ,\\
& \big[A^{(\pm) I}_{0},A^{(\pm) J}_{0}\big]=\varepsilon^{IJK}A^{(\pm) K}_{0}\ , && \big[A^{(\pm) I}_{0},G^{a}_{\pm1/2}\big]=\alpha^{(\pm) I}_{ab}\,G^{b}_{\pm1/2}\ ,\\
& \big\{G^{a}_{\pm1/2},G^{b}_{\pm1/2}\big\}=2\delta^{ab}L_{\pm1}\ , && \big\{G^{a}_{+1/2},G^{b}_{-1/2}\big\}=2\delta^{ab}L_0-2\big(\alpha^{(+)I}_{ab}\,A^{(+)I}_{0}+\alpha^{(-)I}_{ab}\,A^{(-)I}_{0}\big)\ .
\end{aligned}
\end{equation}
Note that in the global supersymmetry algebra realized away from the conformal point, only the vector-like $\mathfrak{su}(2)^{(+)}$ is generally realized (see for example \cite{Aharony:1999dw}).

The off-shell formalisms we will work with make the vector-like combination of the $\lsu(2)^{(+)}$ R-symmetry manifest.   As we will review in Appendix~\ref{SOfour}, the following complex linear combinations of supersymmetry generators will transform as doublets under $\lsu(2)^{(+)}$:
\begin{eqnarray}
& & G^{-}_{\pm1/2}=G^{1}_{\pm1/2}+iG^{2}_{\pm1/2}\ , \\
& & G^{+}_{\pm1/2}=G^{3}_{\pm1/2}-iG^{4}_{\pm1/2}
\end{eqnarray}
We denote the complex conjugates of these as ${\bar G}^{\pm}_{\alpha}$.  In this basis, the supersymmetry algebra (\ref{finitealg}) becomes:
\begin{equation}\label{Gpm}
\begin{aligned}
& \big\{G^{\pm}_{+1/2},G^{\pm}_{-1/2}\big\}=0\ , && \big\{\bar{G}^{\pm}_{+1/2},\bar{G}^{\pm}_{-1/2}\big\}=0\ ,\\
&  \big\{G^{\pm}_{+1/2},G^{\mp}_{-1/2}\big\}=\pm 2iA^{(-)+}_{0}\ , && \big\{\bar{G}^{\pm}_{+1/2},\bar{G}^{\mp}_{-1/2}\big\}=\mp2iA^{(-)-}_{0}\ ,\\
& \big\{G^{+}_{\pm1/2},\bar{G}^{+}_{\pm1/2}\big\}=4L_{\pm1}\ ,&& \big\{G^{-}_{\pm1/2},\bar{G}^{-}_{\pm1/2}\big\}=4L_{\pm1}\ ,\\
& \big\{G^{\pm}_{+1/2},\bar{G}^{\mp}_{-1/2}\big\}=-2iA^{(+)\mp}_{0}\ , && \big\{G^{\pm}_{-1/2},\bar{G}^{\mp}_{+1/2}\big\}=2iA^{(+)\mp}_{0}\ ,\\
& \big\{G^{\pm}_{+1/2},\bar{G}^{\pm}_{-1/2}\big\}=4L_0\mp2i(A^{(+)3}_{0}\mp A^{(-)3}_{0})\ ,&& \big\{G^{\pm}_{-1/2},\bar{G}^{\pm}_{+1/2}\big\}=4L_0\pm2i(A^{(+)3}_{0}\mp A^{(-)3}_{0})\ ,\\
\end{aligned}
\end{equation}
where we have defined the raising and lowering operators of the $\mathfrak{su}(2)$ part as
\begin{equation}
A^{(\pm)+}_{0}=A^{(\pm)1}_{0}+iA^{(\pm)2}_{0}\qquad,\qquad A^{(\pm)-}_{0}=A^{(\pm)1}_{0}-iA^{(\pm)2}_{0}\ .
\end{equation}

\section{${\cal N}=(4,4)$ gravity multiplet}\label{GM}

The off-shell ${\cal N}=4$ Poincar\'e supergravity in two dimensions has the following field content \cite{Gates:1988tn,Gates:1988ey}:
\begin{equation}\label{fcontent}
(e_{\mu}^{\ a},\psi_{\mu,i},\sA_{\mu}^{I},g,h,b)\ ,\qquad i=1,2\ ,\ I=1,2,3\ ,
\end{equation}
where $e_{\mu}^{\ a}$, $\psi_{\mu,i}$, $\sA_{\mu}^{I}$, $g$, $h$, and $b$ are the vielbein, the two Dirac gravitinos, an $SU(2)$-triplet one-form gauge field, an auxiliary real scalar, an auxiliary real scalar, and an auxiliary complex scalar field respectively. This can be derived by various routes \cite{Gates:1988ey}; one route is to start with conformal supergravity, add a conformal scalar multiplet, and fix the gauge freedom of local conformal transformations.  Note that this multiplet makes a single $SU(2)$ R-symmetry explicit; the corresponding Lie algebra is the vector-like combination of the $\mathfrak{su}(2)^{(+)}$ algebra discussed in the previous section.  It would be interesting to consider the case that the R-symmetry of the full large $\CN=4$ algebra is explicit after fixing gauge to arrive at Poincar\'e supergravity. 

The transformation laws of the graviton and the two gravitinos are given by\footnote{We are using the notation in \cite{Gates:1988ey}; a summary of our conventions and a list of useful identities used throughout this paper can be found in appendix~\ref{Notation}.}
\begin{equation}\label{epsiSU}
\begin{aligned}
&\delta e_{\mu}^{\ a}=-\frac{1}{2}\bar{\bar{\epsilon}}^{i}\gamma^a\psi_{\mu,i}+\frac{1}{2}\bar{\bar{\psi}}^{i}_{\mu}\gamma^a\epsilon_{i}\ ,\\
&\delta\psi_{\mu,i}=\sD_{\mu}\epsilon_{i}-\gamma_{\mu}\eta_{i}\ .
\end{aligned}
\end{equation}
where
\begin{equation}\label{sfD}
\begin{aligned}
&\sD_{\mu}\epsilon_{i}=\partial_{\mu}\epsilon_{i}+\frac{1}{4}\omega_{\mu}^{\ ab}\,\gamma_{ab}\,\epsilon_{i}-i\sA_{\mu}^{I}(\sigma^{I})_{i}^{\ j}\,\epsilon_{j} \equiv \del_{\mu} \eps_i - i\sA_{\mu}^{I}(\sigma^{I})_{i}^{\ j}\,\epsilon_{j}\ ,\\
&\eta_{i}=\frac{i}{4}h\,\gamma_{3}\,\epsilon_{i}-\frac{1}{4}g\,\epsilon_{i}-\frac{i}{2}b\,\gamma_{3}\,\epsilon_{i}^{*}\ ,
\end{aligned}
\end{equation}
and $\sigma^I$ are the Pauli matrices. The gauge covariant derivative acting on the conjugate spinor is:
\begin{equation}
	\sD_{\mu} \bbe^i = \d_{\mu} \bbe^i - \frac{1}{4} \omega_{\mu}{}^{ab} \bbe^i \gamma_{ab}
		+ i \sA^I_{\mu} (\sigma^I)^i{}_j \bbe^j \equiv \del_{\mu} \bbe^i + i \sA^I_{\mu} (\sigma^I)^i{}_j \bbe^j
\end{equation}

The transformations and field contents listed above are for the theory in Lorentzian signature.  Once we have this picture, the Euclidean version proceeds by:
\begin{itemize}
\item Promoting real fields to complex fields.
\item Promoting complex conjugate fields such as $b,b^*$ to independent complex fields. This means that we also take $\bbe$ to be independent of $\epsilon$.
\end{itemize}
Upon coupling the background supergravity fields to matter, the resulting action will not in general be real.

Following \cite{Festuccia:2011ws,Dumitrescu:2012ha}, in order to decouple gravity from the theory and to obtain a consistent rigid supersymmetric theory in a curved background, we set the gravitinos and their supersymmetry variations to zero: this ensures that the 2d metric and the background superfields are invariant under supersymmetry. When we ultimately couple the supergravity multiplet to matter, the metric and the auxiliary fields in the gravity multiplet will appear in the supersymmetry variations of the matter fields, leading to a consistent realization of supersymmetry on a curved background.

Setting $\delta\psi_{\mu,i}=0$ in (\ref{epsiSU}), we arrive at the following constraint on the Dirac $SU(2)$-doublet spinors of the curved background    
\begin{equation}\label{SUSYD}
\sD_{\mu}\epsilon_{i}=\gamma_{\mu}\,\eta_{i}\ .
\end{equation}
Rewriting (\ref{SUSYD}) in its explicit form, we find the following conditions on the $\epsilon^i$:
\begin{equation}\label{SUSYcon}
\begin{aligned}
&\nabla_{\mu}\epsilon_{i}=\frac{i}{4}h\gamma_{\mu}\gamma_{3}\epsilon_{i}-\frac{1}{4}g\gamma_{\mu}\epsilon_{i}-\frac{i}{2}b\gamma_{\mu}\gamma_{3}\epsilon_{i}^{*}+i\sA_{\mu}^{I}(\sigma^{I})_{i}^{j}\epsilon_{j}\ ,\\
&\nabla_{\mu}\bbe^{i}=-\frac{i}{4}h\bbe^{i}\gamma_{3}\gamma_{\mu}+\frac{1}{4}g\bbe^{i}\gamma_{\mu}+\frac{i}{2}b^{*}\bar{\epsilon}^{i}\gamma_{3}\gamma_{\mu}-i\sA_{\mu}^{I}(\sigma^{I})_{j}^{i}\bbe^{j}\ .\\
\end{aligned}
\end{equation}
were the auxiliary fields $h$, $g$, and $b$ acquire complex values. One difference from the $\CN=(2,2)$ case is that when $b \neq 0$, the right hand side of the first (second) line contains a term proportional to $\epsilon^*$ ($\epsilon$). A second difference is the appearance of the auxiliary $SU(2)$ gauge field. 

\section{Consistency conditions for ${\cal N}=(4,4)$ supersymmetry}\label{AdSUSY}
\setcounter{equation}{0}

As in \cite{Festuccia:2011ws,Dumitrescu:2012ha,Klare:2012gn,Closset:2014pda}, the existence of spinors satisfying (\ref{SUSYcon}) places constraints on the 2d background $\Sigma$. We will focus on the case that maximal, global ${\cal N} = (4,4)$ supersymmetry is preserved.  This means there are two complex solutions $\{\epsilon^i, \lambda^i\}$ to (\ref{SUSYcon})\ for each $i$. In flat space with no background fields, we could take 
\begin{equation}
	\epsilon^i = \left(\begin{array}{l} \epsilon^{i}_{+} \\ 0 \end{array}\right)\ ; \ \ \ \lambda^i = \left(\begin{array}{l} 0 
	\\ \lambda^{i}_{-} \end{array}\right)
\end{equation}
In curved space, the solutions will not generally have definite chirality; but we will assume that at each point on $\Sigma$, $\epsilon,\lambda$ are linearly independent spinors.

For oriented two-dimensional surfaces, the spacetime is automatically complex and K\"ahler. Following \cite{Festuccia:2011ws,Dumitrescu:2012ha}, we can build K\"ahler forms from bilinears of the spinors $\epsilon_i, \lambda_j$, but these will simply reproduce the canonical K\"ahler form (the 2d volume form).  

Next, we can build a set of complex Killing vectors $\xi^{A,\mu}$ using $\epsilon_i$, $\lambda_i$:
\begin{eqnarray}
	\xi^{1,\mu} & = & \bbe^i \gamma^{\mu} \epsilon_i\nonumber\\
	\xi^{2,\mu} & = & \bbe^i \gamma^{\mu} \lambda_i \nonumber\\
	\xi^{3,\mu} & = & \bbl^i \gamma^{\mu} \epsilon_i \nonumber\\
	\xi^{4,\mu} & = & \bbl^i \gamma^{\mu} \lambda_i
\end{eqnarray}
There is no guarantee that these are independent.  The supersymmetry conditions imply the Killing vector equation
\begin{equation}
\nabla^{\mu}\xi^{A,\nu}+\nabla^{\nu}\xi^{A,\mu}-g^{\mu\nu}\nabla_{\sigma}\xi^{A,\sigma}=0\ .
\end{equation}

Finally, we get constraints on the auxiliary fields by studying the commutator of covariant derivatives acting on the spinors
\begin{eqnarray}
	[\del_{\mu},\del_{\nu}] \epsilon_i & = & \frac{1}{4}R_{\mu\nu}^{\,\,\,\,\,\,\,ab}\gamma_{ab}\epsilon_{i}
	=\frac{1}{4} {\cal R}(x)\varepsilon_{\mu\nu}\gamma_{3}\epsilon_{i} \nonumber\\
	& = & \frac{i}{4} \veps_{\mu\nu} \gamma^{\rho}\left(\d_{\rho} h + i \d_{\rho} g \gamma_3\right)\epsilon_i  - \frac{i}{2} \veps_{\mu\nu} \gamma^{\rho} \d_{\rho} b\ \epsilon^*_i \nonumber\\
	& & \ \ \ \ \ - \frac{1}{8} \veps_{\mu\nu} \left(h^2 + g^2 + 4 b b^*\right) \gamma_3 \eps_i + i \sF_{\mu\nu}^I \sigma^I{}_i{}^j \eps_j
\end{eqnarray}
where
\be
	\sF^I_{\mu\nu} = \d_{\mu} \sA^I_{\nu} - \d_{\nu} \sA^I_{\mu} + 2 \veps^{IJK} \sA^J_{\mu} \sA^K_{\nu}
\ee
is the $SU(2)$-covariant field strength for $\CA$, and:
\begin{eqnarray}
	[\del_{\mu},\del_{\nu}] \bbe^i & = & - \frac{1}{4} \bbe^i \gamma_3 {\cal R}(x)\veps_{\mu\nu} \nonumber\\
	& = & \frac{i}{4} \veps_{\mu\nu} \bbe^i \gamma^{\rho} \left(\d_{\rho} h - i \d_{\rho} g \gamma_3\right) -
		\frac{i}{2}\veps_{\mu\nu} {\bar\eps}^i \gamma^{\rho}\d_{\rho} b^*\nonumber\\
		& & \ \ \ \ \ + \frac{1}{8} \veps_{\mu\nu} \bbe^i \gamma_3 \left(h^2 + g^2 + 4 b b^*\right) - i \bbe^j \sigma^I_j{}^i\sF^I_{\mu\nu} 
\end{eqnarray}
We have the same equations if we substitute $\eps^i  \to \lambda^i$.  Now if $\epsilon^i,\lambda^i$ are linearly independent spinors at each point on $\Sigma$, and we are working in Euclidean space so that $\bbe^i$ is independent of $\epsilon_i$ (that is, not built from the complex conjugate spinor), the integrability conditions are:
\begin{enumerate}
\item $h$, $g$, and $b$ are all constant.
\item $\sF_{\mu\nu}^I = 0$.
\item $\displaystyle R(x) = - \frac{1}{2} \big(h^2 + g^2 + 4 b b^*\big)$
\end{enumerate}
On a compact two-dimensional surface, the flat connection corresponds to Wilson lines of the manifest $SU(2)$ R-symmetry.  As in \cite{Closset:2014pda}, nontrivial Wilson lines will break supersymmetry, so we demand that $\CA^I_{\mu}$ is pure gauge. Note that the first and third condition requires constant curvature, and for spaces of constant positive curvature, we must take $h$ or $g$ to be imaginary, or let $b,b^*$ be independent.

For our later use of constructing an invariant action on a curved space, we need to calculate the Laplacian of the Dirac spinor $\epsilon_{i}$ satisfying the supersymmetry condition (\ref{SUSYcon}). Using (\ref{SUSYcon}), it is straightforward to calculate the action of the Laplacian operator on Dirac spinor doublet $\epsilon_i$
\begin{equation}\label{ddeps}
\sD_{\mu}\sD^{\mu}\epsilon_i=\frac{1}{2}\Big(\frac{h^2}{4}+\frac{g^2}{4}+bb^*\Big)\epsilon_i\ .
\end{equation}

\section{${\cal N}=(4,4)$ vector multiplet}\label{VM}
\setcounter{equation}{0}
In this section, we will construct the supersymmetry transformations and invariant action for an Abelian vector multiplet propagating on a curved two-manifold. We will find that these transformations in general realize the ``large" ${\cal N}=(4,4)$ supersymmetry algebra acting on vector multiplets.

Before considering the case of the curved space, we will briefly summarize the ${\cal N}=(4,4)$ vector multiplet story in flat space, following \cite{Gates:1988ey}.   

\subsection{Flat space}
${\cal N}=4$ vector multiplets in two dimensions have an off-shell formulation with a finite number of auxiliary fields; the multiplets and action can be obtained via dimensional reduction from ${\cal N}=2$ theories in four dimensions  \cite{Grimm:1977xp,Gates:1988ey}. The off-shell multiplet has 16 degrees of freedom, consisting of the following fields:
\begin{equation}\label{vecmul}
(A,B,C,C^{*},V_{\mu},d^{I},\chi_{i},\bbc^{i})\ ,\qquad i=1,2\ ,\ I=1,2,3\ ,
\end{equation}
in which $A$ and $B$ are real propagating scalar fields, $C$ is a complex propagating scalar field, $V_{\mu}$ is a $U(1)$ gauge field, $d^{I}$ are an $SU(2)$ triplet of auxiliary scalar fields, and $\chi_{i}$ are $SU(2)$-doublet Dirac spinors. We will find it useful to combine $A$, $B$ into a complex field $U = A + i B$.

The rigid supersymmetry transformation laws of the fields of the vector multiplet are given by \cite{Gates:1988ey}
\begin{equation} \label{rigidSUSY}
\begin{aligned}
&\delta C=\bar{\epsilon}^{i}\chi_{i}\quad,\quad \delta C^{*}=-\bar{\bar{\chi}}^{i}\epsilon_{i}^{*}\quad,\quad \delta U=\bbe^{i}P_-\chi_{i}+\bbc^{i}P_-\epsilon_{i}\quad,\quad \delta U^*=\bbe^{i}P_+\chi_{i}+\bbc^{i}P_+\epsilon_{i}\ ,\\
&\delta \chi_{i}=\frac{1}{2}\partial_{\mu}C\gamma^{\mu}\epsilon^{*}_{i}-\frac{1}{2}\partial_{\mu}U P_-\gamma^{\mu}\epsilon_{i}-\frac{1}{2}\partial_{\mu}U^*P_+\gamma^{\mu}\epsilon_{i}-\frac{i}{4}F\gamma_{3}\epsilon_{i}-\frac{1}{2}d^{I}(\sigma^{I})_{i}^{\, j}\epsilon_{j}\ , \\
&\delta\bar{\bar{\chi}}^i=\frac{1}{2}\partial_{\mu}C^*\bar{\epsilon}^i\gamma^{\mu}+\frac{1}{2}\partial_{\mu}U\,\bbe^i\gamma^{\mu}P_- + \frac{1}{2}\partial_{\mu}U^*\,\bbe^i\gamma^{\mu}P_+-\frac{i}{4}F\bbe^i\gamma_3-\frac{1}{2}d^I (\sigma^I)^{i}_{j}\bbe^j\ ,\\
&\delta V_{\mu}=\frac{i}{2}(\bbe^{i}\gamma_{\mu}\chi_{i}+\bbc^{i}\gamma_{\mu}\epsilon_{i})\ ,\\
&\delta d^{I}=\frac{1}{2}(\sigma^{I})_{j}^{i}\,\bbe^{j}\gamma^{\mu}\partial_{\mu}\chi_{i}-\frac{1}{2}(\sigma^{I})_{j}^{i}\,\partial_{\mu}\bbc^{j}\gamma^{\mu}\epsilon_{i}\ .
\end{aligned}
\end{equation}
wherfe $P_{\pm} = \half (1 \pm \gamma_3)$.
In (\ref{rigidSUSY}), $F$ is defined as $F=\varepsilon^{\mu\nu}F_{\mu\nu}$ where $F_{\mu\nu}$ is the field strength associated with the abelian vector field $V_{\mu}$. The invariant Lagrangian which is preserved by supersymmetry transformations  (\ref{rigidSUSY}) is \cite{Gates:1988ey}: 
\begin{equation}\label{flatrigidLag}
{\cal L}= \half \partial_{\mu}U\,\partial^{\mu}U^*+\frac{1}{2}\partial_{\mu}C\,\partial^{\mu}C^{*}+\frac{1}{4}F_{\mu\nu}F^{\mu\nu}-\frac{1}{2}d^{I}d^{I}-\bar{\bar{\chi}}^{i}\gamma^{\mu}\partial_{\mu}\chi_{i}\ .
\end{equation}
and is invariant up to total derivatives.

\subsection{Curved space}

To find the transformation laws in the curved space, we replace ordinary derivatives with the appropriate covariant derivatives. We define the two following $SU(2)$-doublet covariant derivatives 
\begin{equation}\label{Dcov}
(\sD_{\mu})_{i}^{j}=\nabla_{\mu}\delta^{j}_{i}-i\sA^{I}_{\mu}(\sigma^{I})_{i}^{j}\quad,\quad (\bar{\sD}_{\mu})_{i}^{j}=\nabla_{\mu}\delta^{j}_{i}+i\sA^{I}_{\mu}(\sigma^{I})_{i}^{j}\ ,
\end{equation}
where $\delta^{j}_{i}$ is the Kronecker delta symbol. In analogy with (\ref{rigidSUSY}), we take the supersymmetry transformation rules for the components of the vector multiplet to be:
\begin{equation}\label{Curverule}
\begin{aligned}
&\delta C=\bar{\epsilon}^{i}\chi_{i}\quad,\quad \delta C^{*}=-\bar{\bar{\chi}}^{i}\epsilon_{i}^{*}\quad,\quad \delta U=\bbe^{i}P_-\chi_{i}+\bbc^{i}P_-\epsilon_{i}\quad,\quad \delta U^*=\bbe^{i}P_+\chi_{i}+\bbc^{i}P_+\epsilon_{i}\ ,\\
&\delta \chi_{i}=\frac{1}{2}(\sD_{\mu})_{i}^{j}(C\gamma^{\mu}\epsilon_{j}^{*})-\frac{1}{2}(\sD_{\mu})_{i}^{j}(U P_- \gamma^{\mu}\epsilon_{j}) - \frac{1}{2}(\sD_{\mu})_{i}^{j}(U^*P_+\gamma^{\mu}\epsilon_{j})-\frac{i}{4}F\gamma_{3}\epsilon_{i}-\frac{1}{2}d^{I}(\sigma^{I})_{i}^{\, j}\epsilon_{j}\ ,\\
&\delta \bbc^{i}=\frac{1}{2}(\bar{\sD}_{\mu})_{j}^{i}(C^{*}\bar{\epsilon}^{j}\gamma^{\mu})+\frac{1}{2}(\bar{\sD}_{\mu})_{j}^{i}(U\bbe^{j}\gamma^{\mu}P_-) + \frac{1}{2}(\bar{\sD}_{\mu})_{j}^{i}(U^*\bbe^{j}\gamma^{\mu}P_+)-\frac{i}{4}F\bbe^{i}\gamma_{3}-\frac{1}{2}d^{I}(\sigma^{I})_{j}^{\, i}\bbe^{j}\ ,\\
&\delta V_{\mu}=\frac{i}{2}(\bbe^{i}\gamma_{\mu}\chi_{i}+\bbc^{i}\gamma_{\mu}\epsilon_{i})\ ,\\
&\delta d^{I}=\frac{1}{2}(\sigma^{I})_{j}^{i}\,\sD_{\mu}(\bbe^{j}\gamma^{\mu}\chi_{i})-\frac{1}{2}(\sigma^{I})_{j}^{i}\,\sD_{\mu}(\bbc^{j}\gamma^{\mu}\epsilon_{i})\ .
\end{aligned}
\end{equation}
The following action is then invariant under (\ref{Curverule}):
\begin{equation}\label{CurvedrigidLag}
\begin{aligned}
\cL=&\frac{1}{2}\nabla_{\mu}U\,\nabla^{\mu}U^*+\frac{1}{2}\nabla_{\mu}C\,\nabla^{\mu}C^{*}-\bar{\bar{\chi}}^{i}\gamma^{\mu}(\sD_{\mu})_{i}^{j}\chi_{j}+\frac{i}{2}h\,\bbc^{i}\gamma_{3}\chi_{i}-\frac{1}{2}g\,\bbc^{i}\chi_{i}\\
&+\frac{1}{2}d^{I}d^{I}+\frac{1}{4}\Big(\frac{h^2}{4}+\frac{g^2}{4}+bb^{*}\Big)UU^*+\frac{1}{4}\Big(\frac{h^2}{4}+\frac{g^2}{4}+bb^{*}\Big)CC^{*}\\
&-\frac{1}{4}\Big(F+\frac{1}{2}(h-ig)U+\frac{1}{2}(h+ig)U^*+b^{*}C+b\,C^{*}\Big)^{2}\ ,
\end{aligned}
\end{equation}
if in addition we impose the constraint $\sF^I_{\mu\nu} = 0$ which follows from the integrability conditions in section \ref{AdSUSY}.

Given the above supersymmetry transformations, we can show that the commutator of these transformations close on the large ${\cal N}=(4,4)$ algebra.  A direct computation gives the first set as:
\begin{equation}\label{vecalgebra2}
\begin{aligned}
&\big[\delta_{\epsilon},\delta_{\bbl}\big]U={\cal L}_{\xi}U+\rho\, U-\beta_{+}\, U+\beta_{-}\,U\ ,\\
&\big[\delta_{\epsilon},\delta_{\bbl}\big]U^*={\cal L}_{\xi}U^*+\rho\, U^*+\beta_{+}\, U^*-\beta_{-}\,U^*\ ,\\
&\big[\delta_{\epsilon},\delta_{\bbl}\big]C={\cal L}_{\xi}C+\rho\, C-\beta_{+}\, C-\beta_{-}\, C\ ,\\
&\big[\delta_{\epsilon},\delta_{\bbl}\big]C^{*}={\cal L}_{\xi}C^*+\rho\, C^*+\beta_{+}\, C^*+\beta_{-}\, C^*\ ,\\
&\big[\delta_{\epsilon},\delta_{\bbl}\big]V^{\mu}=({\cal L}_{\xi}^{\sA}V)^{\mu}+\nabla^{\mu}\Lambda\\
&\big[\delta_{\epsilon},\delta_{\bbl}\big]\chi_{i}={\cal L}_{\xi}^{\sA} \chi_i+\frac{3}{2}\rho\,\chi_i-\beta_{-}\,\chi_{i,L}-\beta_{+}\,\chi_{i,R}\\
&\qquad-\frac{1}{4}(\bbl^j\gamma^{\mu}\gamma^{\nu}\sD_{\mu}\epsilon_i)\gamma_{\nu}\chi_j+\frac{1}{4}(\bbl^j\gamma^{\mu}\gamma^{\nu}\sD_{\mu}\epsilon_j)\gamma_{\nu}\chi_i \ ,\\
&\big[\delta_{\epsilon},\delta_{\bbl}\big]\bbc^{i}={\cal L}_{\xi}^{\sA}\bbc^i+\frac{3}{2}\rho\,\bbc^i+\beta_{-}\,\bbc^{i}_{L}+\beta_{+}\,\bbc^{i}_{R} \\
&\qquad-\frac{1}{4}(\sD_{\mu}\bbl^i\gamma^{\nu}\gamma^{\mu}\epsilon_j)\bbc^j\gamma_{\nu}+\frac{1}{4}(\sD_{\mu}\bbl^j\gamma^{\nu}\gamma^{\mu}\epsilon_j)\bbc^i\gamma_{\nu}\ ,\\
&\big[\delta_{\epsilon},\delta_{\bbl}\big]d^{I}={\cal L}_{\xi}d^I+2\rho\,d^I\\
&\qquad+\frac{1}{4}(\sigma^I)^{i}_{j}\sD_{\mu}\Big(\sD_{\nu}\bbl^j\gamma^{\mu}\gamma^{\nu}P_+ \epsilon_i U + \sD_{\nu}\bbl^j\gamma^{\mu}\gamma^{\nu}P_- \epsilon_i U^*\\ 
&\qquad\hspace{2.5cm}- \bbl^jP_+ \gamma^{\nu}\gamma^{\mu}\sD_{\nu}\epsilon_i U -  \bbl^jP_- \gamma^{\nu}\gamma^{\mu}\sD_{\nu}\epsilon_i U^*\Big)\ .
\end{aligned}
\end{equation}
where 
\begin{equation}\label{defvecalg}
\begin{aligned}
&\chi_{i,L}=P_{-}\chi_i\ , &&\chi_{i,R}=P_{+}\chi_i \ ,\\
&\xi^{\mu}=-\frac{1}{2}\bbl^{i}\gamma^{\mu}\epsilon_{i}\ , &&\rho=-\frac{1}{4}\nabla_{\mu}\big(\bbl^i\gamma^{\mu}\epsilon_i\big)=\frac{1}{2}\nabla_{\mu}\xi^{\mu}\ ,\\
&\beta_{\pm}=\frac{1}{4}\big(\sD_{\mu}\bbl^iP_{\pm}\gamma^{\mu}\epsilon_i-\bbl^i P_{\pm}\gamma^{\mu}\sD_{\mu}\epsilon_i\big)\ , &&\Lambda = - \frac{i}{2} \left(U\, \bbl^i P_+ \eps_i  +U^*\, \bbl^i P_- \eps_i \right)\\
\end{aligned}
\end{equation}
and the action of the Lie derivatives acting on the field components are:
\begin{equation}
\begin{aligned}
&{\cal L}_{\xi}\left(C,C^*,U,U^*\right)=\xi^{\mu}\partial_{\mu}\left(C,C^*,U,U^*\right)\ ,\\
&{\cal L}_{\xi}^{\sA}V={\cal L}_{\xi}V-d^{\sA}(\iota_{\xi}\sA)=\xi^{\mu}F_{\mu\nu}dx^{\nu}\ ,\\
&{\cal L}_{\xi}^{\sA}\chi_i=\xi^{\mu}\sD_{\mu}\chi_i+\frac{1}{4}(\nabla_{\mu}\xi_{\nu})\gamma^{\mu\nu}\chi_i\ ,\\
&{\cal L}_{\xi}^{\sA}\bbc^i=\xi^{\mu}\sD_{\mu}\bbc^i-\frac{1}{4}(\nabla_{\mu}\xi_{\nu})\bbc^i\gamma^{\mu\nu}\ ,\\
&{\cal L}_{\xi}d^I=\xi^{\mu}(\partial_{\mu}d^I-\varepsilon^{IJK}{\sA}^{J}_{\mu}d^{K})\ .\\
\end{aligned}
\end{equation}
Furthermore, the second lines of $\big[\delta_{\epsilon},\delta_{\bbl}\big]V^{\mu}$, $\big[\delta_{\epsilon},\delta_{\bbl}\big]\chi_i$, $\big[\delta_{\epsilon},\delta_{\bbl}\big]\bbc^i$,  and $\big[\delta_{\epsilon},\delta_{\bbl}\big]d^{I}$ vanish, using the supersymmetry conditions (\ref{SUSYcon}) and the fact that $\gamma^{\mu}\gamma^{\nu}\gamma_{\mu}=0$. 

The next set of commutators are:
\begin{equation}\label{commut}
\begin{aligned}
&\big[\delta_{\epsilon},\delta_{\lambda}\big]U=\kappa_{+}\, C^*\\
&\big[\delta_{\epsilon},\delta_{\lambda}\big]U^*=\kappa_{-}\, C^*\\
&\big[\delta_{\epsilon},\delta_{\lambda}\big]C=\kappa_{-}U+\kappa_{+}U^* \\
&\big[\delta_{\epsilon},\delta_{\lambda}\big]C^*=0\\
&\big[\delta_{\epsilon},\delta_{\lambda}\big]V^{\mu}=\nabla^{\mu}\big(\upsilon\,C^*\big)\\
&\big[\delta_{\epsilon},\delta_{\lambda}\big]\chi_i=-\kappa_{-}\chi^{*}_{i,L}-\kappa_{+}\chi^{*}_{i,R}\\
&\qquad -\frac{1}{4}\big(\sD_{\mu}\bar{\lambda}^j\gamma^{\nu}\gamma^{\mu}\epsilon_i\big)\gamma_{\nu}\chi_{j}^{*}+\frac{1}{4}\big(\sD_{\mu}\bar{\epsilon}^j\gamma^{\nu}\gamma^{\mu}\lambda_i\big)\gamma_{\nu}\chi_{j}^{*} \\
&\big[\delta_{\epsilon},\delta_{\lambda}\big]\bbc^i=0\\
&\big[\delta_{\epsilon},\delta_{\lambda}\big]d^I=-\frac{1}{4}(\sigma^I)^{i}_{j}\nabla_{\nu}C^*\nabla_{\mu}\big(\bar{\epsilon}^j\gamma^{\nu}\gamma^{\mu}\lambda_i-\bar{\lambda}^j\gamma^{\nu}\gamma^{\mu}\epsilon_i\big)\\
&\hspace{1.9cm}-\frac{1}{4}(\sigma^I)^{i}_{j}\nabla_{\mu}\Big(\big(\sD_{\nu}\bar{\epsilon}^j\gamma^{\nu}\gamma^{\mu}\lambda_i-\sD_{\nu}\bar{\lambda}^j\gamma^{\nu}\gamma^{\mu}\epsilon_i\big)C^*\Big)
\end{aligned}
\end{equation}
where
\be
\begin{aligned}
	& \kappa_{\pm}=-\frac{1}{2}\big(\sD_{\mu}\bar{\epsilon}^iP_{\pm}\gamma^{\mu}\lambda_i-\bar{\epsilon}^iP_{\pm}\gamma^{\mu}\sD_{\mu}\lambda_i\big)\\
	& \upsilon=\frac{i}{2}\,\bar{\lambda}^i\epsilon_i\\
\end{aligned}
\ee
Equation (\ref{SUSYcon}) guarantees that the second line of the commutator acting on $\chi$ vanishes. The commutator acting on $d^I$ can be reduced to 
\be
	\big[\delta_{\epsilon},\delta_{\lambda}\big]d^I=-\frac{i}{2}(\sigma^I\sigma^J)^{i}_{j}\,\varepsilon^{\mu\nu}{\sF}^{J}_{\mu\nu}\,(\bar{\epsilon}^j\gamma_3\lambda_i)\,C^*
\ee
which vanishes upon imposing the integrability condition $\sF^I_{\mu\nu} = 0$.

Finally, commutators of the conjugate spinors are:
\be\label{conjcommut}
\begin{aligned}
	&\big[\delta_{\bbe},\delta_{\bar{\bar{\lambda}}}\big]U=\kappa_{-}^{*}\,C\\
	&\big[\delta_{\bbe},\delta_{\bar{\bar{\lambda}}}\big]U^*=\kappa_{+}^{*}\,C\\
	& \big[\delta_{\bbe},\delta_{\bar{\bar{\lambda}}}\big]C=0\\
	& \big[\delta_{\bbe},\delta_{\bar{\bar{\lambda}}}\big]C^*=\kappa_{-}^{*}U^*+\kappa_{+}^{*}U\\
	& \big[\delta_{\bbe},\delta_{\bar{\bar{\lambda}}}\big]V^{\mu}=\nabla^{\mu}\big(\upsilon^*\,C\big)\\
	& \big[\delta_{\bbe},\delta_{\bar{\bar{\lambda}}}\big]\chi_i=0\\
	& \big[\delta_{\bbe},\delta_{\bar{\bar{\lambda}}}\big]\bbc^i=-\kappa_{-}^{*}\bar{\chi}^{i}_{L}-\kappa_{+}^{*}\bar{\chi}^{i}_{R}\\
	& \big[\delta_{\bbe},\delta_{\bar{\bar{\lambda}}}\big]d^I=0
\end{aligned}
\ee

While our formalism only makes a single vector $\lsu(2)$ R-symmetry manifest, we claim that the supersymmetry algebra here realizes the ``large" ${\cal N}=(4,4)$ superalgebra, with R-symmetry $\lsu(2)^{(+)}\oplus \lsu(2)^{(-)}$, and $\lsu(2)^{(+)}$ comprising the manifest R-symmetry.  More precisely we find specific actions of the R-symmetry, translations, and dilatations, depending on the background auxiliary fields: the parameters $\rho$, $\beta_{\pm}$, and $\kappa_{\pm}$ depend on the auxiliary fields via the supersymmetry conditions (\ref{SUSYcon}).   

In the case of the fermions, $\lsu(2)^{(+)}$ acts on the doublet index $i$, while $\lsu(2)^{(-)}$ acts on the doublets
\be
	\left(\begin{array}{l} (\chi_1)^* \\ \chi_2 \end{array}\right)\ \qquad \left(\begin{array}{l} - (\chi_2)^* \\ \chi_1 \end{array}\right)
\ee
This can be deduced from Appendix (\ref{SOfour}). 

In the case of the scalars, the left-moving $\lsu(2)^{(-)}$ acts on the doublet
\be
	\left(\begin{array}{l} U \\ C \end{array}\right)
\ee
while the right-moving $\lsu(2)^{(-)}$ acts on the doublet
\be
	\left(\begin{array}{l} U^* \\ C \end{array}\right)
\ee

Given these identifications, Eq. (\ref{Gpm}) is consistent with Eqs. (\ref{vecalgebra2}), (\ref{commut}), and (\ref{conjcommut}), we see that
\begin{itemize}
\item $\xi$ corresponds to translations.
\item $\rho$ corresponds to dilatations.
\item The vectorlike combination of $\lsu(2)^+$ acts on $\chi^i, \bbc^i$ via the term $i \xi^{\mu} \sA_{\mu,i}{}^j$ in $\CL_{\xi}^{\sA}$. 
\item $\beta_{-},\beta_{+}$ corresponds to the left- and right-moving actions of $A_0^{(-)3}$.
\item $\kappa_{-}$ corresponds to the left-moving part of $A^{(-)+}$ and $\kappa_{+}$ to the right-moving part of $A^{(-)+}$.
\item $\kappa_{-}^*$ corresponds to the left-moving part of $A^{(-)-}$ and $\kappa_{+}^*$ to the right-moving action of $A^{(-)-}$.
\item $\Lambda$, $v C^*$, and $v^* C$ are gauge transformations under the complexified gauge group of the abelian vector multiplet.
\end{itemize}

We can view the 2-dimensional ${\cal N}=(4,4)$ supersymmetry as the dimensional reduction of the 6-dimensional ${\cal N}=(1,0)$ theory on a two-manifold \cite{Aharony:1999dw}. The R-symmetry of the 6 dimensional theory reduces to the 2 dimensional $\lsu(2)^{(+)}$ R-symmetry, whereas the $\lso(4)$ rotational symmetry in the 4 extra dimensions becomes the product of the two $\lsu(2)^{(-)}$ symmetries in two dimensions.\footnote{\ We would like to thank our anonymous referee for reminding us of this point.}  The additional components of the 6d gauge field become a quartet of real scalars, transforming as a vector of $\lso(4) = \lsu(2) \oplus \lsu(2)$. The results of Appendix B show how such a vector can be organized into complex doublets of the two $\lsu(2)$ factors. The scalars $U$, $U^*$, $C$, and $C^*$ transform in precisely this fashion.

\section{${\cal N}=(4,4)$ on $S^2$}\label{Sphere}
\setcounter{equation}{0}

A standard test case is to derive the supersymmetry conditions for $S^2$ \cite{Benini:2012ui,Doroud:2012xw,Closset:2014pda}.  Here we consider solutions to the spinor equations for general $h,g,b,b^*$ satisfying the integrability conditions in Section~\ref{AdSUSY}. 

We consider $S^2$ with the metric 
\begin{equation}
	ds^2 = r^2 \left(d\theta^2 + \sin^2 \theta d\phi^2\right)\ ,
\end{equation}
with zweibeins
\begin{equation}
	e^1_{\theta} = r\ , \qquad e^2_{\phi} = r \sin \theta\ ,
\end{equation}
spin connection
\begin{equation}
	\omega_{\phi}{}^1{}_2 = - \cos\theta\ ,
\end{equation}
and Ricci scalar ${\cal R} = \frac{2}{r^2}$. In the Euclidean continuation, the gamma matrices become:
\begin{equation}
	\gamma_1 = \sigma_2,\ \gamma_2 = \sigma_1, \qquad\ \gamma_{12} = - i \sigma_3 = i \gamma_3\ .
\end{equation}

For simplicity, we define
\begin{equation}
\begin{aligned}
	& H = \frac{r}{4} (h + i g)\ ,\\
	& \tH = \frac{r}{4} (h - i g)\ ,\\
	& B = \frac{r}{2} b\ ,
	& \tB = \frac{r}{2} b^*\ .
\end{aligned}
\end{equation}
The curvature constraint from section (\ref{AdSUSY}) becomes $H\tH + B \tB = - \frac{1}{4}$.  In essence, $H,\tH$ are the same as ${\cal H}, {\tilde {\cal H}}$ in \cite{Closset:2014pda}.

We define the spinors $\Psi_{i,\alpha = \pm}$ (where $\alpha$ is the spinor index) as:
\begin{equation}
	\Psi_{i} = \left(\begin{array}{c} \Psi_{i,-} \\ \Psi_{i,+} \end{array}\right)
		= \left(\begin{array}{c} c_i\ \eps_{i,-} - d_i\ \eps^{*}{}_{i,-} \\ 
		-2 i (c_i H + d_i \tB)\ \eps_{i,+} + 2 i (c_i B - d_i \tH)\ \eps^{*}{}_{i,+} \end{array}\right)\ .\\
\end{equation}
With these definitions, and the curvature constraint, the equations for $\chi$ are the standard Killing spinor equations:
\begin{equation}
	\del_{\mu} \Psi_i = \frac{i}{2} \gamma_{\mu} \gamma_3 \Psi_i\ ,
\end{equation}
with solutions 
\begin{equation}
	\Psi_i = C_i\ e^{i\phi/2} \left(\begin{array}{c} \cos\frac{\theta}{2} \\ i \sin \frac{\theta}{2}\end{array}\right)
		+ D_i\ e^{- i \phi/2} \left(\begin{array}{c} \sin\frac{\theta}{2} \\ - i \cos \frac{\theta}{2}\end{array}\right)\ ,
\end{equation}
where $C_i$ and $D_i$ are constants of integration. Note that $\Psi_i$ has the same form as the solutions in \cite{Benini:2012ui}, for each element of the doublet; however, it is a complicated linear combination of the spinors~$\epsilon_i, \epsilon^{*}_{i}$.

\section{Conclusions}\label{Conc}

There is obvious further technical work to be done.  One is to study nonabelian vector multiplets.
Another is to study hypermultiplets on curved backgrounds.  As stated above, hypermultiplets which can be linearly gauged in a straightforward way (that is, the scalars are written as an $SU(2)$ doublet of auxiliary fields) do not have completely off-shell, manifestly $SU(2)_R$ covariant representations with a finite number of auxiliary fields. There is, however, a partially off-shell multiplet in ${\cal N}=2,d=4$ theories -- see for example chapter 12 of \cite{West:1990tg} -- which reduces to a ${\cal N} = (4,4)$ hypermultiplet upon dimensional reduction, for which the supersymmetries close on an off-shell central charge.  There are also other completely off-shell hypermultiplets whose supersymmetry transformations close on a finite number of auxiliary fields \cite{Gates:1988tn,Gates:1995aj}. For instance in one case, the scalars form an $SU(2)$ singlet and $SU(2)$ triplet and so a single multiplet will not transform linearly under gauge transformations.   Nonetheless, these are also worth studying. One could also study off-shell multiplets with infinite numbers of auxiliary fields coupled to supergravity, via harmonic \cite{Galperin:1984av,Galperin:1987em,Galperin:1987ek,Bellucci:2000yx,TartaglinoMazzucchelli:2009jn} or projective \cite{Kuzenko:2008ep,TartaglinoMazzucchelli:2009ip}\ superspace.

An interesting extension of this work would be to study cases analogous to \cite{Closset:2014pda}, with fewer unbroken supercharges, and nonvanishing $\sA^I, \sF^I$.

Finally, it would be interesting to study various localization calculations in this framework; this should be possible at least for noncompact theories, after the fashion of \cite{Harvey:2014nha}.

\vspace{2cm}
\noindent
{\Large{\textbf{Acknowledgements}}}
\vspace{0.4cm}
 
\noindent
We would like to thank Thomas Dumitrescu, S. James Gates, Jim Halverson, Shamit Kachru, Martin Rocek, and David Tong for inspirational and useful discussions and correspondences. Part of this work was performed while A.L. was at the Aspen Center for Physics, which is supported by National Science Foundation grant PHY-1066293. Part of this work was performed while A.L. was visiting the KITP at UC Santa Barbara, supported in part by the National Science Foundation under Grant No. NSF PHY11-25915. A.L. would like to thank both institutions for providing stimulating environments conducive to productive work.  The research of A.L. is supported by DOE grant DE-SC0009987. The research of M.S. is supported by NSF FRG grant DMS 1159049 and NSF grant PHY 1053842. 

\appendix
\setcounter{equation}{0}
\section{Notations and conventions}\label{Notation}
In this section, we provide more details on our notation and conventions. In ${\cal N}=(4,4)$ theory, the parameters of supersymmetry transformation are Dirac spinors which carry an $SU(2)$ index of the R-symmetry. The parameter of supersymmetry, $\epsilon_i$, is an $SU(2)$-doublet Dirac spinor with two spinorial components. The $SU(2)$ R-symmetry indices are raised and lowered by the totally antisymmetric tensor $\epsilon_{ij}$ in the following way 
\begin{equation}
\epsilon^i=\varepsilon^{ij}\epsilon_{j}\qquad,\qquad \epsilon_i=\epsilon^j \varepsilon_{ji}=-\varepsilon_{ij}\epsilon^j\ .
\end{equation}
The $SU(2)$ totally antisymmetric tensor, $\epsilon_{ij}$, fulfills the following properties
\begin{equation}
\varepsilon_{ij}=\varepsilon^{ij}\qquad,\qquad \varepsilon_{ji}\varepsilon^{ik}=-\delta_{j}^{k}\ .
\end{equation}
Moreover, one can relate the product of two $SU(2)$ Kronecker delta to Pauli matrices. The following identity turns out to be very useful  
\begin{equation}\label{delsig}
2\,\delta^{i}_{l}\delta^{j}_{k}=\delta^{i}_{k}\delta^{j}_{l}+(\sigma^{I})^{i}_{k}\,(\sigma^{I})^{j}_{l}\ .
\end{equation}
In the above equation, one performs a sum over the index $I=1,2,3$. Also, one can easily show 
\begin{equation}
A_iB_j-A_jB_i=-\varepsilon_{ij}\,A^k B_k\quad,\quad A^iB^j-A^jB^i=\varepsilon^{ij}\,A^k B_k\ ,
\end{equation}
where $A_i$ and $B_i$ could be both complex numbers, as well as Grassmanians. Complex conjugation raises and lowers the $SU(2)$ R-symmetry indices as follows
\begin{equation}\label{ComCon}
\epsilon^{*i}\equiv(\epsilon_i)^*\qquad,\qquad (\epsilon^i)^*=(\varepsilon^{ij}\epsilon_j)^*=\varepsilon_{ij}\epsilon^{*j}=-\epsilon^{*}_{\,\,\,i}\qquad,\qquad (\epsilon^i)^{**}=-\epsilon^i\ .
\end{equation}
In particular, we notice in (\ref{ComCon}) that after two successive complex conjugations, a minus sign is produced. For spinors, we introduce two types of conjugation: namely the Majorana and the Dirac conjugations. These two spinors are respectively defined as
\begin{equation}\label{epCon}
\bar{\epsilon}^i=-\varepsilon^{ij}\epsilon^{\sf{T}}_{j}\cC\qquad,\qquad \bbe^i=\varepsilon^{ij}(\epsilon_j)^{\dagger}\cC\ ,
\end{equation}
where $\epsilon^{i\sf{T}}$ is the transpose spinor without complex conjugation, whereas $(\epsilon_i)^{\dagger}$ includes complex conjugation operation in addition to transposition. In (\ref{epCon}), $\cC$ is the two dimensional charge conjugation matrix which happens to coincide with the Pauli matrix $\sigma^2$ 
\begin{equation}
\cC=\sigma^2=\left(\begin{matrix}
0 & -i\\
i & 0
\end{matrix}\right)\ .
\end{equation}
In our conventions, the $\gamma$-matrices which are two dimensional representations of the Clifford algebra are matrices with real entries. The two $\gamma$-matrices $\gamma_0$ and $\gamma_1$ together with the helicity matrix $\gamma_3$ are defined by
\begin{equation}\label{gammat}
\gamma_{0}=\left(\begin{matrix}
0 & -1\\ 1& 0\end{matrix}\right)\qquad,\qquad \gamma_{1}=\left(\begin{matrix}
0 & 1\\ 1& 0\end{matrix}\right)\qquad,\qquad \gamma_{3}=\gamma_{0}\gamma_{1}=\left(\begin{matrix}
-1 & 0\\ 0& 1\end{matrix}\right)\ .
\end{equation}
Using the Clifford algebra, and the explicit representation of $\gamma$-matrices in (\ref{gammat}), it is straightforward to derive the following useful properties among product of $\gamma$-matrices
\begin{equation}
\begin{aligned}
&\gamma_{\mu}^{\sf{T}}\cC=-\cC\gamma_{\mu}\ ,\\
&\{\cC,\gamma_{3}\}=0\ ,\\
&\gamma^{\mu}\gamma^{\nu}=g^{\mu\nu}+\gamma^{\mu\nu}=g^{\mu\nu}+\varepsilon^{\mu\nu}\gamma_{3}\ ,\\
&\gamma^{\mu}\gamma_{\nu}\gamma_{\mu}=0\ ,\\
&\varepsilon^{\mu\nu}\gamma_{\nu}=\gamma_3\gamma^{\mu}\ .
\end{aligned}
\end{equation}
Fierz identity for both the Majorana and Dirac conjugations takes the same form and in our notation is expressed in the following form
\begin{equation}\label{fierz}
\begin{aligned}
&(\bar{\epsilon} \lambda)\chi=+\frac{1}{2}\Big((\bar{\epsilon}\chi)\lambda+(\bar{\epsilon}\gamma_{3}\chi)\gamma_{3}\lambda+(\bar{\epsilon}\gamma^{\mu}\chi)\gamma_{\mu}\lambda\Big)\ ,\\
&(\bbe \lambda)\chi=-\frac{1}{2}\Big((\bbe\chi)\lambda+(\bbe\gamma_{3}\chi)\gamma_{3}\lambda+(\bbe\gamma^{\mu}\chi)\gamma_{\mu}\lambda\Big)\ ,
\end{aligned}
\end{equation}
in which $\epsilon$, $\lambda$, and $\chi$ are all Dirac spinors in two dimensions. From the Fierz identity (\ref{fierz}), we can derive the following useful relations between the product of spinor bilinears 
\begin{equation}
\begin{aligned}
&(\bar{\lambda}\gamma^{\mu}\lambda)(\bar{\epsilon}\gamma_{\mu}\epsilon)+(\bar{\lambda}\gamma^{\mu}\gamma_3\lambda)(\bar{\epsilon}\gamma_{\mu}\gamma_3\epsilon)=0\ ,\\
&(\bar{\epsilon}\gamma^{\mu}\epsilon)(\bar{\epsilon}\gamma_{\mu}\epsilon)+(\bar{\epsilon}\gamma^{\mu}\gamma_3\epsilon)(\bar{\epsilon}\gamma_{\mu}\gamma_3\epsilon)=0\ ,\\
&(\bar{\epsilon}\epsilon)(\bar{\epsilon}\epsilon)+(\bar{\epsilon}\gamma_3\epsilon)(\bar{\epsilon}\gamma_3\epsilon)=0\ ,\\
&(\bar{\bar{\lambda}}\gamma^{\mu}\lambda)(\bbe\gamma_{\mu}\epsilon)+(\bar{\bar{\lambda}}\gamma^{\mu}\gamma_3\lambda)(\bbe\gamma_{\mu}\gamma_3\epsilon)=0\ ,\\
&(\bbe\gamma^{\mu}\epsilon)(\bbe\gamma_{\mu}\epsilon)+(\bbe\gamma^{\mu}\gamma_3\epsilon)(\bbe\gamma_{\mu}\gamma_3\epsilon)=0\ ,\\
&(\bbe\epsilon)(\bbe\epsilon)+(\bbe\gamma_3\epsilon)(\bbe\gamma_3\epsilon)=0\ .
\end{aligned}
\end{equation}

\section{$\lso(4)$ and $\lsu(2)\times \lsu(2)$}\label{SOfour}

\setcounter{equation}{0}

In this work we move between the $SO(4)$ and $SU(2)\times SU(2)$ representations of the R-symmetry of the``large" $\CN=4$ supersymmetry algebra. We provide a few details of this equivalence here in case it is useful to the reader.

The defining representation of the $\lso(4)$ algebra consists of real $4\times 4$ antisymmetric matrices, which we label as
\begin{equation}
	M^{\alpha\beta}_{ab} = \frac{1}{2}\big(\delta_a^{\alpha}\delta_b^{\beta} - \delta_a^{\beta}\delta_b^{\alpha}\big)\ .
\end{equation}
Here $(a,b) = 1,\ldots,4$ label the matrix indices, and $\alpha,\beta = (1,\dots,4)$ label elements of the algebra.  $M^{\alpha\beta}$ generates an infinitesimal rotation of a four-vector in the $\alpha-\beta$ plane.

We can split the orthogonal axes of $\R^4$ in three different ways, corresponding to orthogonal pairs of two-planes.  The two $\lsu(2)$ subalgebras correspond simultaneous rotations along both pairs of planes, with differing relative signs of the rotation.  More specifically, we can rewrite the basis elements 
$\alpha^{(\pm)}_{ab}$ as:
\begin{equation}
\begin{aligned}
	&\alpha^{(\pm) 1} & = & M^{23} \pm M^{14}\ ,\\
	&\alpha^{(\pm) 2} & = & M^{31} \pm M^{24} \ ,\\
	&\alpha^{(\pm) 3} & = & M^{12} \pm M^{34}\ .
\end{aligned}
\end{equation}

Given a vector $V^a$ in $\R^4$, the linear combination
\begin{equation}
	\left(\begin{array}{l} V^{(+)+} \\ V^{(+)-} \end{array}\right) 
	= \left( \begin{array}{l}  V^1 + i V^2 \\ V^3 - i V^4 \end{array}\right)\ ,
\end{equation}
is a doublet under $\lsu(2)^{(+)}$, as is the complex conjugate. Similarly, the linear combination
\begin{equation}
	\left(\begin{array}{l} V^{(-)+} \\ V^{(-)-} \end{array}\right) 
	= \left( \begin{array}{l}  V^1 + i V^2 \\ V^3 + i V^4 \end{array}\right)\ ,
\end{equation}
is a doublet under $\lsu(2)^{(-)}$, as is its complex conjugate.

\eject
\bibliographystyle{utphys}
\bibliography{localizerefs}

\end{document}